\documentclass[referee]{aa}

\usepackage{amsmath}
\usepackage{amssymb}
\usepackage{latexsym}
\usepackage{graphicx}
\usepackage[varg]{txfonts}
\usepackage{natbib}
\usepackage{color}
\bibpunct{(}{)}{;}{a}{}{,} 

\usepackage{pdflscape}
\usepackage{afterpage}

\begin{document}
\title{Model fitting of kink waves in the solar atmosphere: Gaussian damping and time-dependence}

\author{R. J. Morton \& K. Mooroogen} 

\institute{Department of Mathematics \& Information Sciences, Northumbria University, Newcastle Upon Tyne,
NE1 8ST, UK
}

\abstract{}
{Observations of the solar atmosphere have shown that magnetohydrodynamic waves are ubiquitous throughout. Improvements in instrumentation and the 
techniques used for measurement of the waves now enables subtleties of competing theoretical models to be compared with the observed
waves behaviour. Some studies have 
already begun to undertake this process. However, the techniques employed for model comparison have generally been unsuitable and can 
lead to erroneous conclusions about the best model. The aim here is to introduce some 
robust statistical techniques for model comparison to the solar waves community, drawing on the experiences from other areas of 
astrophysics. In the process, we also aim to investigate the physics of coronal loop oscillations. }
{The methodology exploits least-squares fitting to compare models to observational data. We demonstrate that the residuals between 
the model and observations contain significant information about the ability for the model to describe the observations, and show how
they can be assessed using various statistical tests. In particular we discuss the Kolmogorov-Smirnoff one and two 
sample tests, as well as the runs test. We also highlight the importance of including any observational trend line in the model-fitting process.}
{To demonstrate the methodology, an observation of an oscillating coronal loop undergoing standing kink motion is used. The model 
comparison techniques provide evidence that a Gaussian damping profile provides a better description of the observed wave attenuation
than the often used exponential profile. This supports previous analysis from \cite{PASetal2016}. Further, we use the model 
comparison to provide evidence of time-dependent wave properties of a kink oscillation, attributing the behaviour to the 
thermodynamic evolution of the local plasma.}
{}

\keywords{Sun: Corona, Waves, magnetohydrodynamics (MHD), Sun: oscillations}

\date{Received /Accepted}

\titlerunning{Model Fitting}
\authorrunning{Morton \& Mooroogen}

\maketitle

\section{Introduction}

The Sun's atmosphere is known to be replete with magnetohydrodynamic (MHD) wave phenomena and current instrumentation has enabled 
the measurement of the waves and their properties. In particular, the transverse kink motions of magnetic structures in the chromosphere 
and corona has received a great deal of attention, mainly due to their suitability for energy transfer and also their ability 
for providing diagnostics of the local plasma environment via coronal or solar magnetoseismology.

Current instrumentation has demonstrated the capabilities required to accurately measure the motions of the fine-scale magnetic 
structure, with high spatial and temporal resolution and high 
signal-to-noise levels. Further, the almost continuous coverage of {Solar Dynamic Observatory} (SDO), supplemented with 
numerous ground-based data sets and {Hinode} data, has led to 
the existence of a large catalogue of wave events. The result of these fortuitous circumstances means that there is plenty of high 
quality data on kink waves, which can be used for probing the physics 
of the waves and the local plasma. However, these resources have 
yet to be exploited effectively, with only simple modelling of individual observed wave events. By modelling, we refer to using the 
observational data to test theoretical ideas of wave behaviour by fitting an 
expected model. 

In a recent study by \cite{PASetal2016}, the authors attempt to exploit the catalogue of events to do just this. Observations of 
standing kink waves are utilised to test different models of the damping 
profile of the oscillatory motion. The authors attempt to determine whether the observed oscillatory motion can be best described by 
an exponential or Gaussian damping profile, after recent analytical 
studies of the resonant absorption of kink modes suggested that a Gaussian profile is more 
suitable (\citealp{HOOetal2013}; \citealp{PASetal2013}). However, the method used 
by the authors to obtain the time-series data for model fitting 
and the technique used for a model comparison are unfortunately inadequate. The technique used by the authors ignores known problems 
with, and associated uncertainties of, the $\chi^2$ statistic. 
As such, there is the potential to end up with erroneous values for model parameters, as well as underestimating the associated 
uncertainties. More importantly, this methodology is not suitable for distinguishing 
between the two non-linear models. 

{ The study of \cite{PASetal2016} is not the first to fit non-exponential profiles to damped kink oscillations. \cite{DEMetal2002b} 
measured the decay in amplitude of a kink oscillation from wavelet analysis of a displacement time-series. A least-squares fit of a
function of the form $\exp\left(-\epsilon t^n\right)$ is performed for three cases, finding values of $n=\{1.79, 2.83, 0.42\}$.  While the fit
visually appear to describe the data well, the uncertainties are not displayed. The amplitude
spectrum obtained from Fourier techniques is not a consistent statistical estimator of the true amplitude spectrum of the oscillatory
signal, largely because of the noise present in the signal, but also the choice of windowing functions and mother wavelets can cause problems
owing to for example, spectral leakage. A discussion of the significance of Fourier-based methods takes us away from the central
theme of this manuscript, so we do not expand on this any further.}

{\cite{VERetal2004} also investigate whether the damping profile of kink waves observed in a coronal loop arcade is 
exponential. A trend is subtracted from a displacement time-series before a weighted non-linear least-squares fit
of damped sinusoids is performed. The damping term has the same form as \cite{DEMetal2002b}, but the values of $n$ were 
chosen, namely $n=1,2,3$. The authors state they cannot distinguish between the different damping profiles from the
fits, although it is not evident how this comparison was made. }\\

In the following, we outline a methodology for comparing models to observations, providing statistical estimators that can be used to 
assess the suitability of different models. The pitfalls of the 
methodology used in previous studies that perform least-squares fitting are highlighted. We demonstrate that, for a particular example of an
oscillating coronal loop, the evidence suggests that the Gaussian damping is the better 
profile to explain the observed attenuation. Further, the model comparison also reveals that the oscillatory signal contains signatures 
for the dynamic evolution of the background plasma, signified by 
time-dependent behaviour of the wave period.

\begin{figure*}[!tp]
\centering
\includegraphics[scale=0.9, clip=true, viewport=0.cm 0.cm 21.cm 7.5cm]{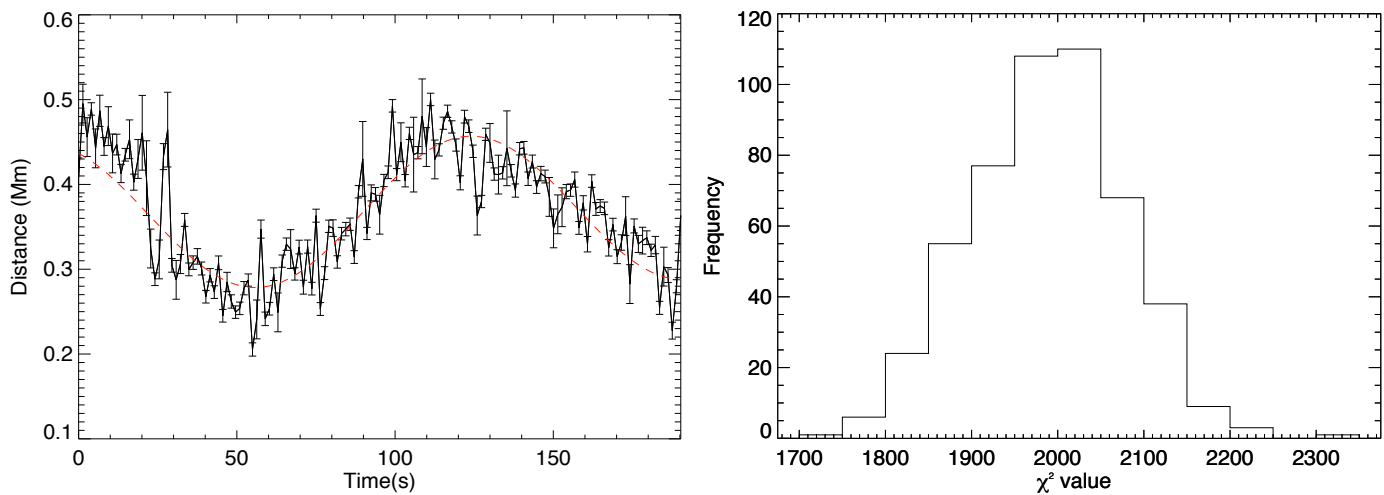} 
\caption{Transverse displacement of a quiescent chromospheric fibril. The time-series of 141 data points represent the location of 
the centre of the fibrils'
cross-sectional flux profile, with the uncertainties in the location given by the error bars (left panel). The overplotted red dashed 
line shows
the best fit of the model. The distribution of the $\chi^2$ statistic for 500 realisations of the random noise in the time-series (right
panel).   }\label{fig:ts_chrom}
\end{figure*}

\section{Observations and measurement}
Here, two different data sets are used for the purpose of demonstrating the techniques for modelling. The first is a H$\alpha$ data set from
the 5 May 2013 taken at the \textit{Swedish Solar Telescope} (SST) with the CRisp Imaging SpectroPolarimeter (CRISP - \citealp{SCHetal2008}). 
Details of the data reduction technique are given in \cite{KURetal2015}. The cadence of the observations is 1.34~s and spatial sampling is 
0".0592~pixel$^{-1}$.

The second data set is $171$~{\AA} data from SDO \textit{Atmospheric Imaging Assembly} (AIA) recorded
on 7 January 2013. The loop analysed is Event 43 Loop 4 using the notation from the catalogue of standing kink oscillations identified in \cite{
GODetal2016}.
The time-distance diagram is created from a location that is high along the loop leg, relatively close to the loop apex.
The data was prepared using the standard AIA solarsoft packages. The cadence of the observations is 12~s and spatial sampling is 0".6~pixel$^{-1
}$. 

The technique used to measure the waves from these data sets involves the fitting of a Gaussian function to the cross-sectional flux profile of the 
structures in each time-frame, weighting the fits with uncertainties in intensity. {The fitting of a Gaussian is an established and common technique 
for estimating the location of the central position of an oscillating structure (e.g., \citealp{ASCetal2002}; \citealp{VERetal2004}).} Full details 
of the technique used here can be found in \cite{MOR2014} and the code is freely available (\citealp{MOR_NUWT}, Zenodo, doi:10.5281/zenodo.49563). 
The estimates for the uncertainties are obtained from: (i) the equations for noise in AIA data (e.g., \citealp{YUANAK2012}); 
(ii) image processing techniques to isolate and model the noise in the SST data (\citealp{MOOetal2016}).

For both data sets, an estimate of residual motion between frames is required. To assess this, suitable regions of the full field of view (FOV) 
are identified
and cross-correlation analysis is performed giving displacement vectors. The displacement vectors contain signatures of the
long-term, large-scale movement of the chromospheric/coronal features due to underlying photospheric flows and/or solar rotation. As such, we 
de-trend the displacement vectors with high order polynomials to 
remove such motions, leaving essentially a time-series of residual frame-to-frame motion that is thought to be present due to either 
atmospheric seeing (SST) or space-craft jitter (SDO). This is 
performed for a number of regions in each data set, and the root mean square of each residual motion time-series is calculated. The RMS values 
of motion in the $x$ and $y$ directions are summed 
in quadrature and averaged between the different regions to provide a single value for each data set, which we take to represent the residual 
motion. This value is added in quadrature to the
uncertainties of all Gaussian centroid positions measured from the  cross-sectional flux profiles.

As we highlight in the next section, this technique probably under estimates the residual motion in the ground-based data. This is likely 
due to the processing of the data via the MOMFBD and alignment
techniques (e.g., \citealp{VANNetal2005}) that `de-stretches and warps' relatively small subsections of the data to remove the influence of 
atmospheric seeing conditions, which can vary between different parts of an 
image. As such, regions of worse seeing may result in increased difficulties in aligning data between frames and lead to larger 
uncertainties on the relative positions of magnetic features between frames.

\begin{figure}[!tp]
\centering
\includegraphics[scale=0.55, clip=true, viewport=1.cm 3.8cm 18.cm 9.cm]{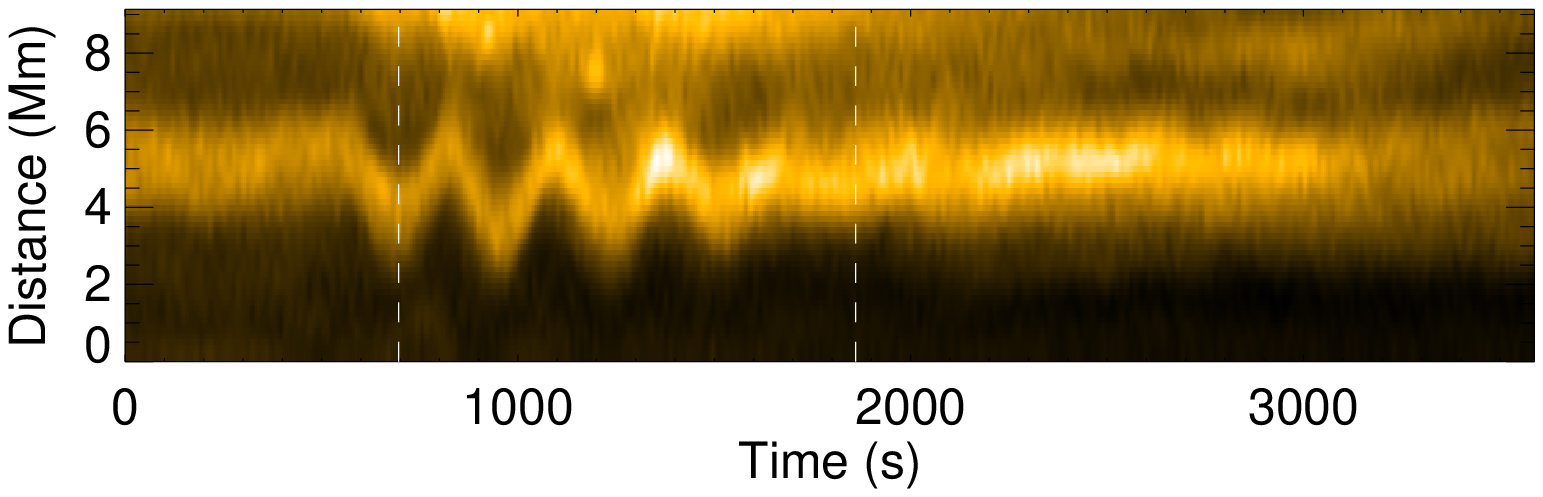} 
\includegraphics[scale=0.55, clip=true, viewport=1.0cm 0.cm 18.cm 12.2cm]{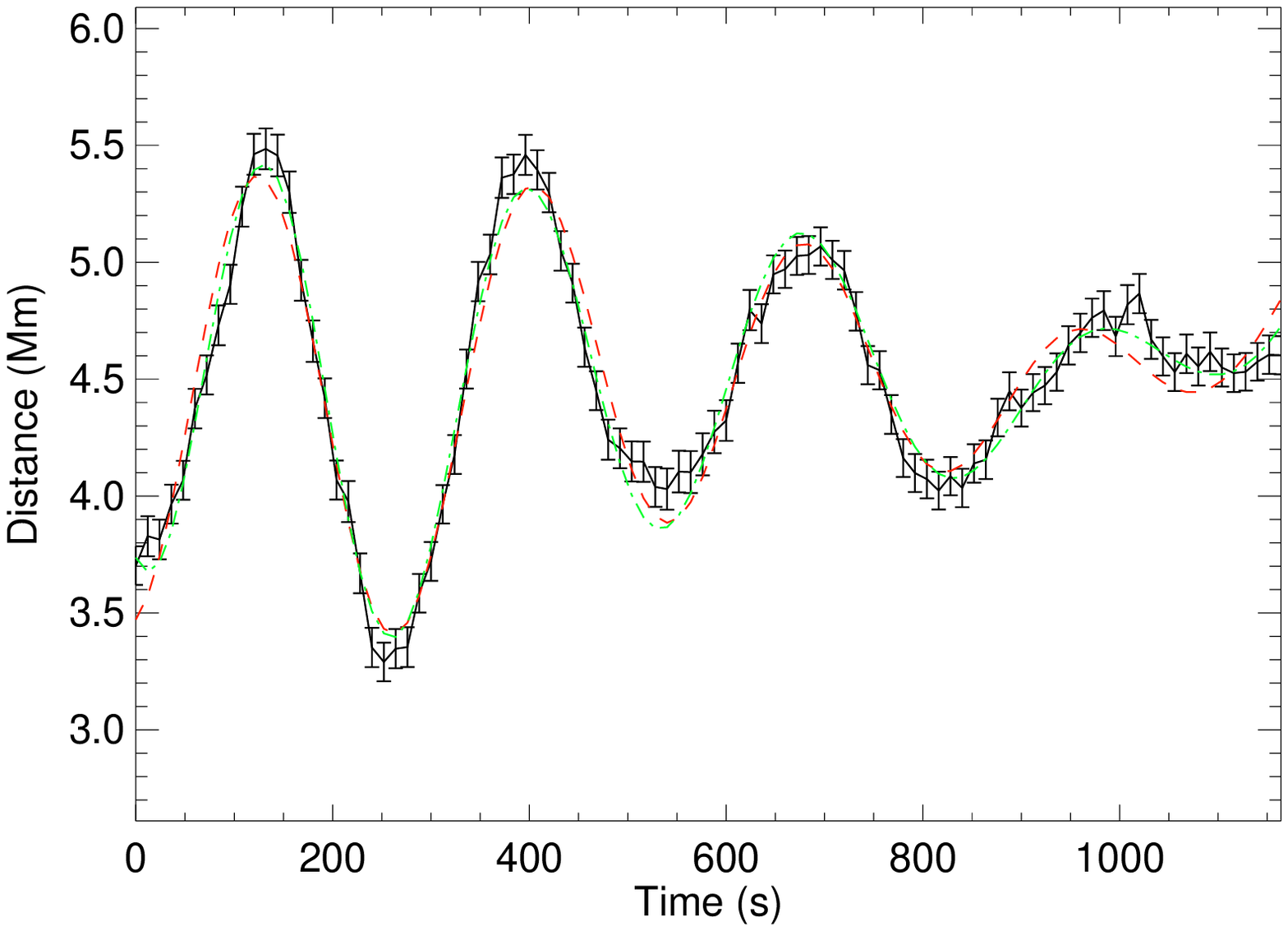} 
\caption{Model fitting a kink oscillation of a coronal loop. The top panel shows the time-distance diagram for the coronal loop oscillation, where the time shown is in seconds is from 06:29:59~UT. The bottom panel shows the time-series of 97 data points representing the location of the centre of the coronal loop's
cross-sectional flux profile used for the model fitting (solid black curve) with the error bars showing the standard deviation of uncertainties. The time shown is in seconds is from 06:41:23~UT and the corresponding section of the time-distance diagram is highlighted by the dashed white lines in the top panel. The red dashed curve shows the best fit quartic Gaussian model and the green dash-dot curve shows the time-dependant model, $f_3$.}\label{fig:slit}
\end{figure}

\section{Model comparison}
Here, we utilise least-squares minimisation to fit a model to the observed data. The technique involves comparing some model $f$ with parameters $a$ to $N$ data values, $y(t_i)$, with errors $\sigma_i$ measured at $t_i$ by essentially minimising the $\chi^2$, given
by 
\begin{equation}\label{eq:chi}
\chi^2=\sum_{i=1}^{N}\left\{ \frac{y(t_i)-f(t_i;a) } {\sigma_i}\right\}^2.
\end{equation}

A standard approach to model comparison is to find a single value of the reduced $\chi^2$, i.e., $\chi^2_\nu=\chi^2/DOF$, where $DOF$ is the number of degrees of freedom, and compare the values of $\chi^2_\nu$ for the different models. However, as discussed in \cite{ANDetal2010}, the use and definition of $\chi^2_\nu$ is subject to various problems if the model is non-linear and, as such, is not a suitable quantity for model comparison. Further, it is clear from Eq.~\ref{eq:chi} that the value of the $\chi^2$ is subject to uncertainty if the data contains random noise, i.e., random noise can lead to variations in measured values of $y(t_i)$, hence, the value of $\chi^2$ is also dependent upon the data noise. 

\begin{figure*}[!tph]
\centering
\includegraphics[scale=1., clip=true, viewport=0.cm 6.5cm 18.cm 12.7cm]{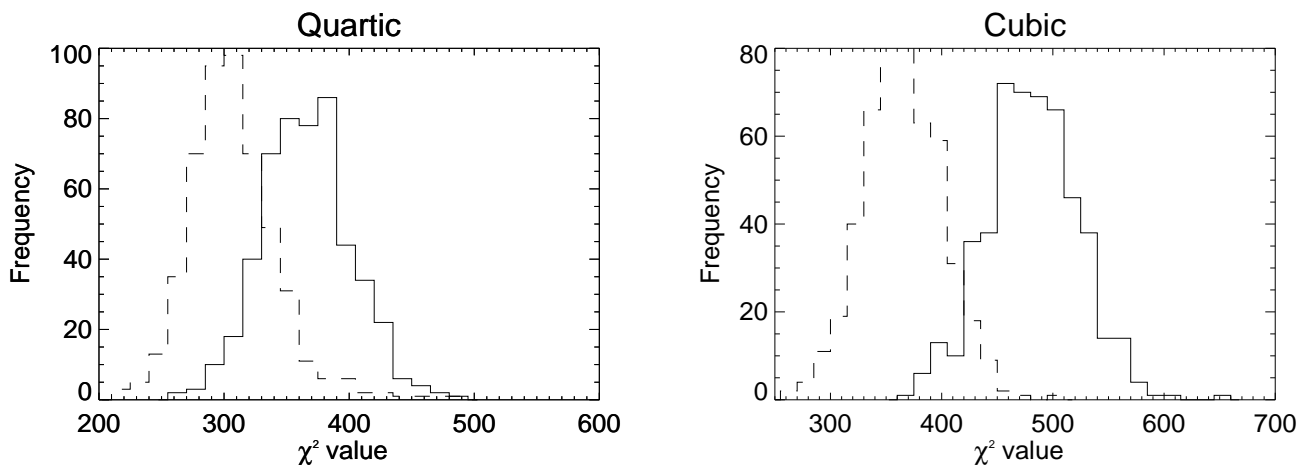} 

\caption{Distribution of the $\chi^2$ statistic for the quartic and cubic models. The solid lines show the distribution for the exponential damping, while the dashed line shows the distribution for the Gaussian damping. }\label{fig:chisq}
\end{figure*}

As a demonstration of this, we take an example of a kink wave time-series measured in the chromosphere. In Figure~\ref{fig:ts_chrom}, the displayed data is the measured displacement of the central 
axis of a fibril in the quiescent chromosphere. This example is chosen because the wave motion shows a relatively simple displacement profile that is very near sinusoidal and we do not have to 
concern ourselves (too much) about extra physics. That data is fit with a simple model of the form,
\begin{equation}
f=a_0+a_1t+a_2\sin(2\pi t/a_3+a_4),
\end{equation}
where $a_n$ are the parameters to be determined. Fitting the data using the Levenberg algorithm (\textit{mpfit} - \citealp{MAR2009}), with the determined uncertainties, provides a $\chi^2$ value of 
$1910$. 

To demonstrate the variability of the $\chi^2$, we use re-sampling to create new versions of the time-series. Each data point, $y(t_i)$, has an associated uncertainty ($\sigma_i$), implying the 
measured value of $y(t_i)$ is the `actual' value plus some deviation due to the random noise, with the value of noise coming from a Gaussian distribution centred on the `actual' value with variance $
\sigma_i^2$. For each data point $y(t_i)$, we generate new values by randomly sampling the noise from a Gaussian distribution with the data points given uncertainty. This process is performed to 
create five hundred representations of the time-series. The model is fit to each representation and the $\chi^2$ is calculated. This process mimics the influence of the random error associated with 
each data point and Figure~\ref{fig:ts_chrom} shows the resulting distribution for the $\chi^2$, which has a mean and standard deviation of $1988\pm90$. Hence, it is clear that quoting a single value 
for $\chi^2$ is not adequate for determining the uncertainty associated with the fitted model.

As mentioned earlier, from examination of the variability of the time-series and the given uncertainties, the estimated values for some of the $\sigma_i$ do not represent the observed 
variance between time-frames particularly well (left panel Figure~\ref{fig:ts_chrom}). There are a number of large excursions, lying at least $5\sigma$ from the fitted model. This is likely due to an 
underestimation of the residual motion of the data between frames in this region. As such, we are likely over-estimating the magnitude of $\chi^2$ associated with this data. It is worth noting that 
under-estimation of $\sigma_i$ leads to a wider distribution of the $\chi^2$, as well as over-estimating its magnitude.

\smallskip
There are a number of well established statistical techniques to overcome the unsuitability of the reduced $\chi^2$ and enabling model comparison to take place. A brief discussion of such techniques 
is given in \cite{ANDetal2010} and we do not detail the various techniques here. By way of example, in the following we perform model comparison for a standing coronal kink 
oscillation, whose measured time-series shows clear indications that detailed physics is required in the model in order to describe the observation.

\subsection{Example: Exponential vs. Gaussian damping}

The first step in model comparison is choosing an appropriate model for the time-series that is based upon physical considerations. In analysing the 
loop oscillations, \cite{PASetal2016} first de-trends 
the oscillations using a spline fit to the maxima and minima of the oscillation. De-trending signals is a popular technique used in time-series 
analysis (e.g., \citealp{VERetal2004, VERetal2010b}; \citealp{VANetal2007}) but there are a number of problems that arise with 
de-trending a series before model fitting (see also, e.g., \citealp{VAU2005} and \citealp{GRUetal2011}, for a similar example with significance 
testing of Fourier components). First, it removes physics 
from the data. While the consequences of de-trending using low order polynomials may be small on the values of fitted parameters, complicated spline 
fits of the data 
are likely to skew results and influence model comparison, since the data can be manipulated in any fashion. On viewing the subtracted 'trends` in \cite
{PASetal2016}, it appears physical effects 
are being removed, with the trend-lines showing oscillatory behaviour. Second, the fitted trend influences the maximum likelihood values for the 
model parameters obtained, hence, the residuals and 
the value of $\chi^2$. Since the goal is to derive accurate estimates for the model parameters and their uncertainties, the removal of the 
trend is counter-productive and likely leads to an 
underestimation of the uncertainty with fitted parameters. As such, the trend should not be removed before model fitting, but should form part of 
the model fitting and the comparison process.

\smallskip
Before performing the model comparison, we select a portion of the time-series to fit (Figure~\ref{fig:slit}). We exclude the beginning of the time-series from 06:29:59~UT to 06:41:23~UT when the wave is excited. If we did not, the model would have to be complicated enough to include the physics related to the excitation of the wave. Since the modelling of the excitation process is not the focus here, the beginning of the time-series is excluded. We also exclude the last section of the time-series from 07:01:11~UT. From here, the displacement amplitude of the wave becomes comparable to the size of the uncertainties, hence this section of the time-series can add no new information to the model fitting.

\begin{figure*}[!tp]
\centering
\includegraphics[scale=1., clip=true, viewport=0.cm 0.cm 18.cm 12.2cm]{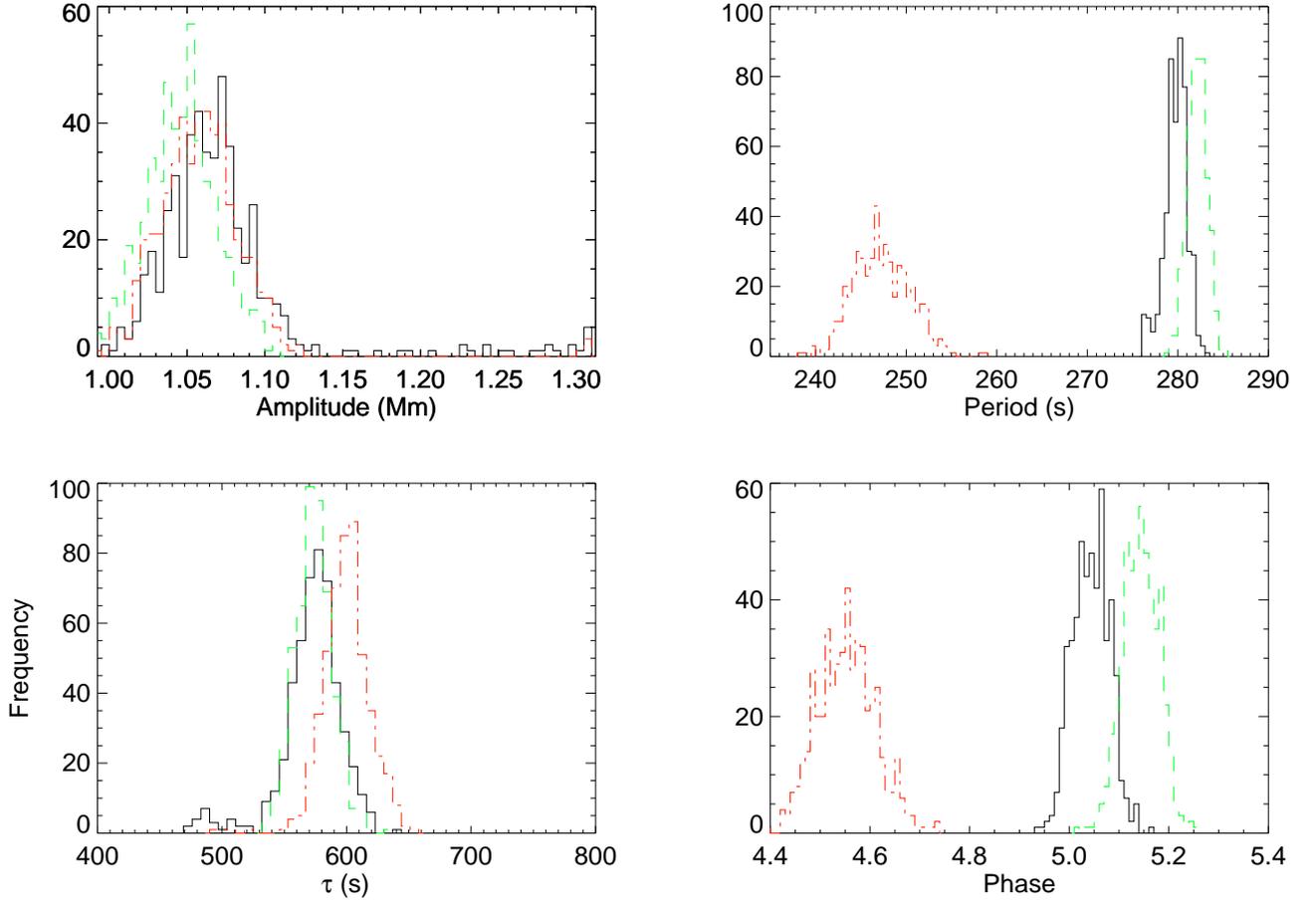} 

\caption{Distributions of fitted parameters from the quartic Gaussian damping (solid black), cubic Gaussian damping (dashed green)
and quartic Gaussian damping with time-dependent period (dash-dot red).  }\label{fig:fit_param}
\end{figure*}

We begin by choosing a low-order polynomial as the base for the model. Coronal loops are likely subject to long time-scale motions from the shuffling and movement of its footpoints by the horizontal flows in the photosphere. An exact physical model for the trend is unknown (e.g., \citealp{ASCetal2002}). It is likely that such motions will be non-linear, hence we opt for a maximum of a fourth-order polynomial to describe any long-term evolution of the loop. In order to demonstrate the influence of the `trend' on the fitted parameters, we also fit a cubic model. 
 
Further, the data shows evidence for damped sinusoidal displacement of the coronal loop, hence a damped sinusoid is required in the model. The damping of kink waves in coronal loops is a well studied phenomenon and resonant absorption is the most promising candidate to explain the observed damping (e.g., \citealp{RUDROB2002}, \citealp{GOOetal2002}, \citealp{HOOetal2013}, \citealp{PASetal2013}). Finally, we would like to assess two models that are associated with resonant absorption, one with a Gaussian damping profile and another with an exponential profile, hence we multiply the sinusoid by one of these damping terms. The two sets of models for comparison are then:
\begin{eqnarray}
f_1 &= a_0\sin(2\pi t/a_1+a_2)\exp(-a_3t)+\nonumber\\
& a_4+ a_5t+a_6t^2+a_7t^3 \mbox{ } (+a_8t^4) , \\
&\nonumber\\
f_2 &= a_0\sin(2\pi t/a_1+a_2)\exp(-a_3t^2)+\nonumber\\
& a_4+ a_5t+a_6t^2+a_7t^3  \mbox{ } (+a_8t^4),
\end{eqnarray}
where $f_1$ is the exponential profile and $f_2$ the Gaussian profile. Here, for the exponential profile $a_3=1/\tau_e$ where $\tau_e$ is the damping time, and for the Gaussian profile $a_3=1/(2\tau_g^2)$  where $\tau_g$ is the damping time.

These models are fit to the data separately. As demonstrated in the previous section, the $\chi^2$ is subject to uncertainties. To assess this, we 
use re-sampling of the uncertainties to obtain distributions for the $\chi^2$ values (Figure~\ref{fig:chisq}) and the fitted parameters of the model 
(some of which are shown in Figure~\ref{fig:fit_param}). The mean values of $\chi^2$ and their standard deviation are given in Table~\ref{tab:meas}. 
The $\chi^2$ values appear to suggest that the Gaussian damping profile performs better then exponential damping profile independent of the trend 
line used, minimising the sum of squared residuals. It is worth highlighting here that if we use a single of $\chi^2$ to judge the quality of fit, 
in a small number of cases it is found that the exponential model out performs the Gaussian, i.e., has a smaller value of  $\chi^2$. If one of these 
realisations of the random error just happened to be the measured values, then relying on a single value of the $\chi^2$ would lead us to wrongly 
conclude the exponential is a better model.

\begin{figure}[!tp]
\centering
\includegraphics[scale=0.56, clip=true, viewport=1.1cm 0.cm 17.2cm 12.cm]{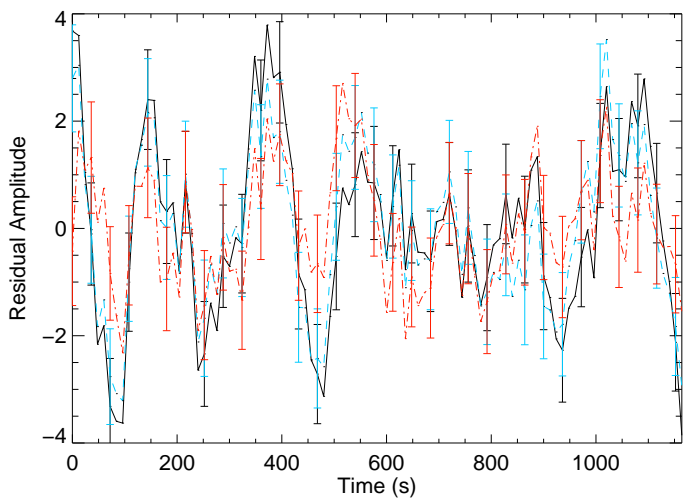} 
\caption{Averaged normalised residuals for the different models with the fourth-order polynomial trend. The black solid line shows
the exponential damping profile, the blue dashed line is the Gaussian damping profile and the red dash-dot is the time-dependent model.
The normalised error bars are also displayed for every third data point. }\label{fig:resid}
\end{figure}

\medskip
However, it is not clear that either model is a particular good approximation to the real data. If the model contains all the physics encapsulated 
in the data, then we should expect that the normalised residuals should be random and distributed normally, i.e., $\mathcal{N}(0,1)$. In Figure~
\ref{fig:resid} we show the average normalised residuals between the  the resampled time-series and models. The residuals are quite clearly not
randomly distributed, with apparent periodic structure present. A non-parametric statistical test called the \textit{runs test for randomness} 
(e.g., \citealp{BARLOW1989}; \citealp{WALJEN}) can be implemented for the residuals. {The test begins by assigning a 0 or 1 to the residuals 
if it lies below or above the mean value respectively. Should the residuals be random and normally distributed, the average number of runs of 0s 
and 1s and the variance for the expected number of runs can be estimated. Comparing this to the observed number of runs forms the hypothesis test.} Perhaps predictably, there is evidence at the 95$\%$ level to 
suggest that the residuals are not randomly distributed.

\begin{figure}[!tp]
\centering
\includegraphics[scale=0.45, clip=true, viewport=0.cm 0.cm 20.cm 11.9cm]{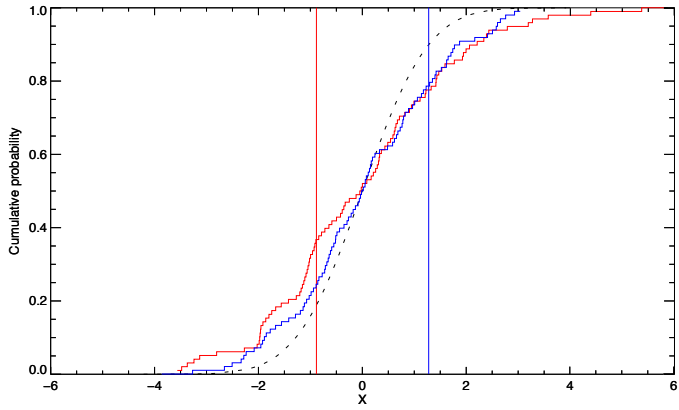} 

\caption{Examples of the KS one sample test. The ECDF of the residuals for the cubic Gaussian damping (red), time-dependent quartic Gaussian damping (blue) and normal CDF (dashed). The
solid vertical lines show the location of the maximum distance between the ECDF and CDF. }\label{fig:ks_ecdf}
\end{figure}

\begin{figure*}[!htp]
\centering
\includegraphics[scale=1., clip=true, viewport=0.cm 6.5cm 18.cm 12.7cm]{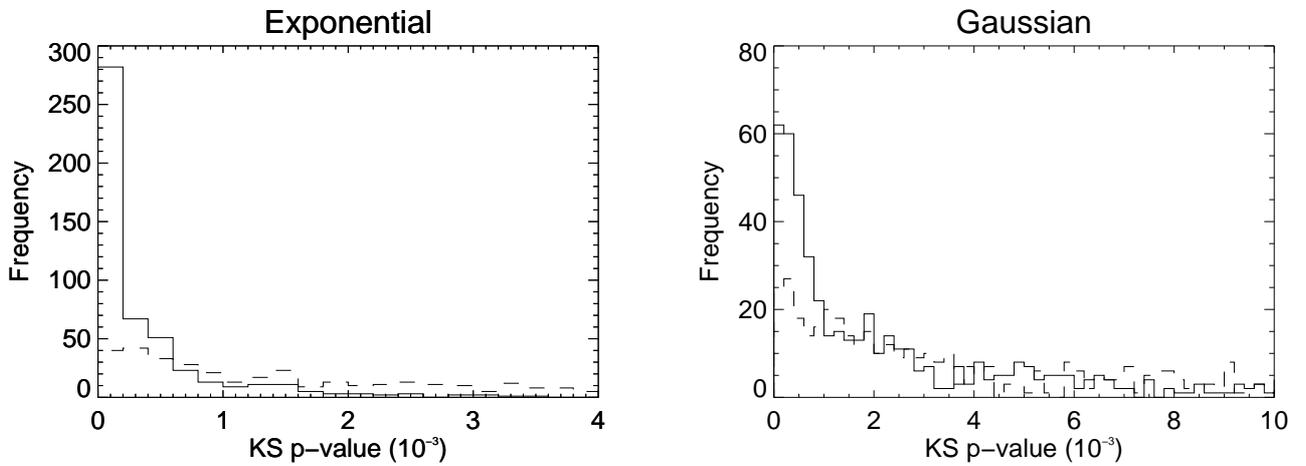} 
\caption{Histogram of p-values from KS test. The solid and dashed lines show the values for the cubic and quartic polynomials, respectively.}\label{fig:ks}
\end{figure*}

\smallskip
In order to quantify the quality of the fitted models, we perform a residuals test that compares the distribution of the normalised residuals to a normal distribution. The model that is favoured by the 
data is the one whose normalised residuals best match the normal distribution. A powerful test to perform this comparison is the non-parametric Kolmogorov-Smirnoff (KS) one sample test (e.g., 
\citealp{BABFEI}, \citealp{WALJEN}), which compares the theoretical cumulative distribution function (CDF) to the sample empirical (E)CDF, examples of which are given in Figure~\ref{fig:ks_ecdf}. 
The test uses the null hypothesis that the sample distribution is from the theoretical CDF and calculates the maximum deviation between the sample and theoretical CDF as the test statistic. The value 
of the maximum deviation can be used to calculate a probability, which essentially provides an estimate of the probability required for the true model to produce a residual equal to the maximum 
deviation. Hence, small \textit{p}-values imply that there is evidence that the null hypothesis is false, i.e., the residuals are not from a normal distribution. In Figure~\ref{fig:ks}, we display the 
\textit{p}-values from the KS test. As with the $\chi^2$, the \textit{p}-value is also subject to uncertainty, so we calculate the value for each of the 500 re-samples. It is clear from the values in Figure~\ref{fig:ks} 
that non of the models used appear to accurately describes the physics in the data, although the Gaussian damping profile is favoured more strongly over the exponential damping profile. The mean 
value of the distributions is given in Table~\ref{tab:meas} and agrees with the impression from the plotted distributions. 

\smallskip
Focusing now on the model parameters obtained from the fitting, we demonstrate the importance of including the trend line in the fitting of the 
model. It is evident from Figure~\ref{fig:chisq} that the $\chi^2$ value depends on the trend, hence, any further tests that may be carried out for 
goodness of fit using the residuals will also be dependent upon the trend used. Further, it is found that whether fitting with a cubic or quartic 
spline influences the values of the model parameters. As an example, it is found that the initial amplitude of the wave is greater using the quartic 
trend rather the cubic (see also Table~\ref{tab:meas}). We can further quantify any differences between the the two distributions of amplitudes from 
the two fitted trends by testing whether are from the same parent distribution. This involves using the non-parametric two-sample KS test to compare 
the two ECDFs, whose null hypothesis is that the two distributions are from the same parent population. The test results provide evidence 
at the $95\%$ level to suggest that the two distributions are from different parent distributions. 

{As an additional test, we performed the least-squares analysis on trend subtracted data. First, a quartic trend is fitted by weighted 
least-squares
to the time-series and the found trend is removed. In line with previous studies that remove the trend first, the uncertainties associated with the 
fitted trend line are not added to the original errors. Performing a least-squares fit of a damped sinusoid to the trend subtracted series, weighted
with the original errors, gives
similar values for parameters but underestimates their uncertainties by $20\%$.} 

Hence, it is clear that not taking into account 
the trend will bias the results of the model fitting and also any further tests of the goodness of fit of the models.

\subsection{Example: Time-dependence}
As indicated by examination of the residuals, it is suggested that the physics encapsulated in the fitted model did not accurately describe the observed oscillation.
There is potentially much physics we could incorporate into the model, but perhaps the most obvious exclusion is
the dynamical behaviour of the background plasma. It is now well documented that a significant proportion of observed coronal loops undergo evolution, continually out of hydrostatic equilibrium, 
exhibiting cycles of heating and cooling where mass is frequently exchanged with the lower solar atmosphere through evaporation and condensation  (e.g., \citealp{ASCetal2000b}, 
\citealp{WINetal2003}, \citealp{ASCTER2008}, \citealp{UGAetal2009}, \citealp{VIAKLI2012}, \citealp{TRIetal2012}, \citealp{MCIetal2012}). Oscillations and waves can exist on this dynamically 
evolving background plasma, and the variation in plasma properties has been demonstrated to alter both the period and amplitude of waves (e.g., \citealp{MORERD2009b}, 
\citealp{MORHOOERD2010}, \citealp{RUD2010}, \citealp{TERetal2011}, \citealp{CHEetal2015}).

\begin{figure}[!tp]
\centering
\includegraphics[scale=0.54, clip=true, viewport=1.cm 0.0cm 18.cm 12.7cm]{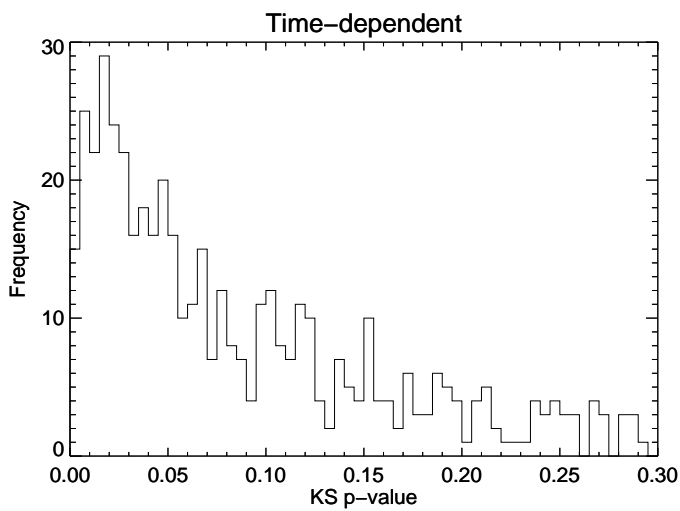} 
\caption{Histogram of p-values from KS test for the time-dependent model.  }\label{fig:ks_tdp}
\end{figure}

It is therefore natural to extend our model to include dynamic effects. The exact form of the time-dependence for a particular physical mechanism is likely to be cumbersome and we do not know which 
dynamical effects are occurring. Hence, a simple approximation for the dynamics is employed, assuming that the period is time-dependent with the form $P_0(1+t/\tau_d)$, where $P_0$ is the initial 
period and $\tau_d$ represents the dynamic time-scale of change. So, we now fit a model of the form,
\begin{eqnarray}
f_3 =& a_0\sin\left(\frac{2\pi t}{a_1(1+a_2t)}+a_3\right)\exp(-a_4t^2)+\nonumber\\
 &a_5+ a_6t+a_7t^2+a_8t^3+a_9t^4.
\end{eqnarray}
The mean values of the parameter distributions from the fitting of the re-sampled data are found and used to define the best fit model, which is 
overplotted in Figure~\ref{fig:slit}. {The value obtained for $a_2=1.36\pm0.14\times10^{-4}$~s$^{-1}$, meaning the period increases from 247~s to 288~s over the course of the time-series.}  Close inspection 
reveals that this model fits the data better than any of the previous models. This is supported by the distribution of the $\chi^2$ statistic, whose 
mean occurs at significantly smaller values than the 
previous models (Table~\ref{tab:meas}). Further, the residual test results have dramatically improved, with the mean value of KS test statistic 
increasing by a factor of 10 compared to the time-independent quartic Gaussian 
profile (Figure~\ref{fig:ks_tdp}). It worth noting that $\sim58\%$ of the re-sampled series have KS test statistic values that fall below the 
critical value at the $95\%$ level, i.e., in these cases there is no 
evidence to suggest that residuals are not normally distributed. 

\begin{figure}[!tp]
\centering
\includegraphics[scale=0.53, clip=true, viewport=0.5cm 0.0cm 17.cm 11.2cm]{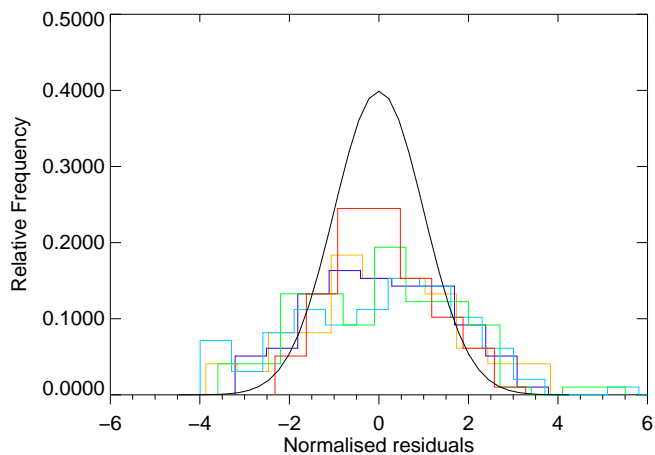} 
\caption{Distribution of normalised residuals for all the fitted models. The black line shows a normal distribution. The legend for the different distributions is as follows: quartic Gaussian (purple) ; quartic exponential (yellow); cubic Gaussian (green) ; cubic exponential (blue); time-dependent (red). }\label{fig:norm_resid}
\end{figure}

A better way to visualise the comparison of residuals is to show the distribution of normalised residuals in relation to normal distribution. In Figure~\ref{fig:norm_resid} the results from all the models 
used are shown. It is clear for the time-independent models that the distribution of residuals is much wider than the normal distribution, with significantly more values greater than 3 standard deviations 
from the mean than expected. The time-dependent model however, has much less extreme values of the residuals and can be seen to be approaching a normal distribution.

\section{Discussion}
The observation of wave phenomenon throughout the solar atmosphere now occurs with regularity, and techniques have been developed to provide high quality measurements of the observed 
waves properties. The quality of these measurements can be utilised to probe the local plasma of the wave guide through seismology (e.g., \citealp{MOR2014}, \citealp{WANetal2015}). Significantly, 
this enables models of oscillatory phenomenon to be tested against observed cases, with the quality of available data enabling us move beyond very basic models, e.g., a simple sinusoid, and probe increasingly complex physics. 
However, in order to make 
decisions about whether the chosen models describe the observed behaviour well, robust techniques for model comparison are required. These techniques give an indication of the probability that 
the observed behaviour will arise, assuming the chosen model is correct. A range of such techniques have already been developed and are used widely in astrophysics, and many other areas. 

Here, we demonstrate the benefit of using these techniques for the modelling of oscillations of fine-scale magnetic structure in the corona. The unsuitability of single value $\chi^2$ and also $\chi^2_\nu$ statistics to 
compare models is highlighted, as discussed in \cite{ANDetal2010}, and we demonstrate how greater confidence in the comparison of competing models
can be achieved by using robust statistical techniques. 
In particular, the techniques were applied to the kink oscillations of a coronal loop. It was found that the amplitude profile of the kink wave 
was best described by a Gaussian damping profile rather than an exponential damping profile. This result had already been suggested in \cite{PASetal2016}, although the methodology used in that 
study to reach this conclusion is based upon comparison of single values of $\chi^2_\nu$, and as such, no confidence intervals where associated with the conclusions. Moreover, in a number of cases, \cite{PASetal2016} suggest the exponential profile is a better fit to the data than a Gaussian profile. These cases could potentially be false results, due to either: (i) the subtraction of a spline profile that distorts the true amplitude envelope; (ii) reliance upon a single value of $\chi^2$ (or more precisely $\chi_\nu^2$); (iii) not including the trend line in the assessment of uncertainties.\\

{
Similar comments can be made regarding previous results that use trend subtraction and then try to estimate other parameters. For example,
\cite{VANetal2007} aim to measure multiple harmonics of a kink oscillation associated with a coronal loop. The displacement time-series 
is trend subtracted and fit with a single frequency sinusoid. The residuals between the trend subtracted data and single frequency model are then analysed by an additional 
least-squares fit of a secondary sinusoidal component. We suggest that the uncertainties associated with the parameters from the fitting of 
these residuals would be
significantly greater than given by the authors. The subtraction of the trend alone is seen to underestimate the uncertainties (Section 3.1), but an 
additional fit to the residuals will only exacerbate this effect on the parameter estimates of the secondary sinusoidal model. To ensure statistical significance,
we suggest a single model incorporating all the physics should be fit to the original time-series. }

{It is then unclear whether the secondary harmonic found in \cite{VANetal2007} is statistically significant. The lack of error bars on the data and residual plots also do not enable a visual assessment of the situation. Applying both the runs test and the KS test
would help elucidate whether any additional structure was contained with the residuals above the uncertainties.
}\\

Further, we extended the Gaussian damping model to include the effects of time-dependence, namely through a simple modification of the period to 
permit it to vary as a function of time. Such a model 
naturally arises when the background plasma that supports the oscillation is subject to thermodynamic evolution, e.g., heating/cooling (cf \citealp{
MORERD2009b}, \citealp{RUD2010,RUD2011}). We note that the amplitude of the
oscillation is also influenced by the time-dependent background plasma, indicating that the amplitude envelope described by the Gaussian damping 
contains the influence of both dynamics and 
attenuation by e.g., resonant absorption. The analysis performed here suggests that the time-dependent model describes the data better than the static 
model. We believe this is evidence of dynamic coronal loop plasma influencing the properties of the oscillation. {A change in periodicity
over the observed displacement time-series has previously been reported by \cite{DEMetal2002b} and \cite{WHIetal2013}.}\\ 

Further work is need to untangle the information that the fitted model provides about the plasma evolution and the 
attenuation due to resonant absorption, and should be complemented with analysis of the plasma and magnetic field, e.g, differential emission measure. Moreover, 
while the time-dependent model performs much better than the time-independent model, it is still clear that something is missing from the current 
analysis (e.g., as demonstrated by the non-random residuals in Figure~\ref{fig:resid} 
and the KS statistic in Figure~\ref{fig:ks_tdp}). This could potentially be due to under-estimates of uncertainty, which 
would imply are model describes all the physics occurring during this event. However, we believe it is more likely that some physics in the data is 
not captured by the model and this is supported by the lack of randomness in the residuals (Figure~\ref{fig:resid}) . It is unknown what this may be 
at present. One possibility 
is the data contains the signature of higher harmonics of the kink wave, which could also have been excited along with the fundamental mode (e.g.,
\citealp{VERetal2004}; \citealp{VANetal2007}; \citealp{DEMBRA2007}; \citealp{OSHetal2007}; 
\citealp{VERERDJES2008}). In order to find answers to some of these questions, an extended study will be required.

\begin{table*}
\caption{Parameter estimates and statistics from the model comparison.}
\centering

\begin{tabular}{lccccc}

\hline\hline\
Model & $\chi^2$ & Mean KS  & Amplitude (km) & Period (s) & $\tau$ (s) \\[0.2ex]
& &test p-value & & & \\
\hline
\\
Cubic Exponential  & 488$\pm$43 & $4\times10^{-4}$ & 1183$\pm$30 & 284$\pm$1& 811$\pm$34\\
Cubic Gaussian  & 369$\pm$36 & 0.003 & 1044$\pm$22 & 282$\pm$1 & $572\pm25$\\
Quartic Exponential  & 371$\pm$36 & 0.005  & 1236$\pm$32 & 281$\pm$1 & 795$\pm$38\\
Quartic Gaussian & 310$\pm$37  & 0.01 & 1070$\pm$48 & 280$\pm$1  & $574\pm15$\\
Time-dependent & 207$\pm31$ & 0.1 & 1061$\pm$30 & 247$\pm$3 & $600\pm19$\\
\hline

\end{tabular}

\label{tab:meas}
\end{table*}

\begin{acknowledgements}
RM is grateful to the Leverhulme Trust for the award of an Early Career 
Fellowship. KM thanks Northumbria University for the award of a PhD studentship. 
The authors acknowledge IDL support provided by STFC. The authors also
thank V. Henriquez for allowing access to the SST data.
\end{acknowledgements}

\bibliographystyle{aa}

\begin{thebibliography}{43}
\expandafter\ifx\csname natexlab\endcsname\relax\def\natexlab#1{#1}\fi

\bibitem[{{Andrae} {et~al.}(2010){Andrae}, {Schulze-Hartung}, \&
  {Melchior}}]{ANDetal2010}
{Andrae}, R., {Schulze-Hartung}, T., \& {Melchior}, P. 2010, ArXiv e-prints

\bibitem[{{Aschwanden} {et~al.}(2002){Aschwanden}, {de Pontieu}, {Schrijver},
  \& {Title}}]{ASCetal2002}
{Aschwanden}, M.~J., {de Pontieu}, B., {Schrijver}, C.~J., \& {Title}, A.~M.
  2002, \solphys, 206, 99

\bibitem[{{Aschwanden} {et~al.}(2000){Aschwanden}, {Nightingale}, \&
  {Alexander}}]{ASCetal2000b}
{Aschwanden}, M.~J., {Nightingale}, R.~W., \& {Alexander}, D. 2000, \apj, 541,
  1059

\bibitem[{{Aschwanden} \& {Terradas}(2008)}]{ASCTER2008}
{Aschwanden}, M.~J. \& {Terradas}, J. 2008, \apjl, 686, L127

\bibitem[{{Babu} \& {Feigelson}(1996)}]{BABFEI}
{Babu}, G.~J. \& {Feigelson}, E.~D. 1996, Astrostatistics (Chapman and Hall)

\bibitem[{{Barlow}(1989)}]{BARLOW1989}
{Barlow}, R. 1989, {Statistics. A guide to the use of statistical methods in
  the physical sciences}

\bibitem[{{Chen} {et~al.}(2014){Chen}, {Li}, {Xia}, {Chen}, \&
  {Yu}}]{CHEetal2015}
{Chen}, S.-X., {Li}, B., {Xia}, L.-D., {Chen}, Y.-J., \& {Yu}, H. 2014,
  \solphys, 289, 1663

\bibitem[{{De Moortel} \& {Brady}(2007)}]{DEMBRA2007}
{De Moortel}, I. \& {Brady}, C.~S. 2007, \apj, 664, 1210

\bibitem[{{De Moortel} {et~al.}(2002){De Moortel}, {Hood}, \&
  {Ireland}}]{DEMetal2002b}
{De Moortel}, I., {Hood}, A.~W., \& {Ireland}, J. 2002, \aap, 381, 311

\bibitem[{{Goddard} {et~al.}(2016){Goddard}, {Nistic{\`o}}, {Nakariakov}, \&
  {Zimovets}}]{GODetal2016}
{Goddard}, C.~R., {Nistic{\`o}}, G., {Nakariakov}, V.~M., \& {Zimovets}, I.~V.
  2016, \aap, 585, A137

\bibitem[{{Goossens} {et~al.}(2002){Goossens}, {Andries}, \&
  {Aschwanden}}]{GOOetal2002}
{Goossens}, M., {Andries}, J., \& {Aschwanden}, M.~J. 2002, \aap, 394, L39

\bibitem[{{Gruber} {et~al.}(2011){Gruber}, {Lachowicz}, {Bissaldi}, {Briggs},
  {Connaughton}, {Greiner}, {van der Horst}, {Kanbach}, {Rau}, {Bhat}, {Diehl},
  {von Kienlin}, {Kippen}, {Meegan}, {Paciesas}, {Preece}, \&
  {Wilson-Hodge}}]{GRUetal2011}
{Gruber}, D., {Lachowicz}, P., {Bissaldi}, E., {et~al.} 2011, \aap, 533, A61

\bibitem[{{Hood} {et~al.}(2013){Hood}, {Ruderman}, {Pascoe}, {De Moortel},
  {Terradas}, \& {Wright}}]{HOOetal2013}
{Hood}, A.~W., {Ruderman}, M., {Pascoe}, D.~J., {et~al.} 2013, \aap, 551, A39

\bibitem[{{Kuridze} {et~al.}(2015){Kuridze}, {Henriques}, {Mathioudakis},
  {Erd{\'e}lyi}, {Zaqarashvili}, {Shelyag}, {Keys}, \& {Keenan}}]{KURetal2015}
{Kuridze}, D., {Henriques}, V., {Mathioudakis}, M., {et~al.} 2015, \apj, 802,
  26

\bibitem[{{Markwardt}(2009)}]{MAR2009}
{Markwardt}, C.~B. 2009, in ASPCS, Vol. 411, Astronomical Data Analysis
  Software and Systems XVIII, ed. D.~A. {Bohlender}, D.~{Durand}, \&
  P.~{Dowler}, 251

\bibitem[{{McIntosh} {et~al.}(2012){McIntosh}, {Tian}, {Sechler}, \& {De
  Pontieu}}]{MCIetal2012}
{McIntosh}, S.~W., {Tian}, H., {Sechler}, M., \& {De Pontieu}, B. 2012, \apj,
  749, 60

\bibitem[{{Mooroogen} {et~al.}(2016){Mooroogen}, {Morton}, \&
  {Henriques}}]{MOOetal2016}
{Mooroogen}, K., {Morton}, R.~J., \& {Henriques}, V. 2016, In prep.

\bibitem[{{Morton}(2014)}]{MOR2014}
{Morton}, R.~J. 2014, \aap, 566, A90

\bibitem[{{Morton} \& {Erd{\'e}lyi}(2009)}]{MORERD2009b}
{Morton}, R.~J. \& {Erd{\'e}lyi}, R. 2009, \apj, 707, 750

\bibitem[{{Morton} {et~al.}(2010){Morton}, {Hood}, \&
  {Erd{\'e}lyi}}]{MORHOOERD2010}
{Morton}, R.~J., {Hood}, A.~W., \& {Erd{\'e}lyi}, R. 2010, \aap, 512, A23+

\bibitem[{Morton {et~al.}(2016)Morton, Mooroogen, \& McLaughlin}]{MOR_NUWT}
Morton, R.~J., Mooroogen, K., \& McLaughlin, J.~A. 2016, {NUWT: Northumbria
  University Wave Tracking (NUWT) code}

\bibitem[{{O'Shea} {et~al.}(2007){O'Shea}, {Srivastava}, {Doyle}, \&
  {Banerjee}}]{OSHetal2007}
{O'Shea}, E., {Srivastava}, A.~K., {Doyle}, J.~G., \& {Banerjee}, D. 2007,
  \apj, 473, L13

\bibitem[{{Pascoe} {et~al.}(2016){Pascoe}, {Goddard}, {Nistic{\`o}},
  {Anfinogentov}, \& {Nakariakov}}]{PASetal2016}
{Pascoe}, D.~J., {Goddard}, C.~R., {Nistic{\`o}}, G., {Anfinogentov}, S., \&
  {Nakariakov}, V.~M. 2016, \aap, 585, L6

\bibitem[{{Pascoe} {et~al.}(2013){Pascoe}, {Hood}, {De Moortel}, \&
  {Wright}}]{PASetal2013}
{Pascoe}, D.~J., {Hood}, A.~W., {De Moortel}, I., \& {Wright}, A.~N. 2013,
  \aap, 551, A40

\bibitem[{{Ruderman}(2010)}]{RUD2010}
{Ruderman}, M.~S. 2010, \solphys, 267, 377

\bibitem[{{Ruderman}(2011)}]{RUD2011}
{Ruderman}, M.~S. 2011, \solphys, 114

\bibitem[{{Ruderman} \& {Roberts}(2002)}]{RUDROB2002}
{Ruderman}, M.~S. \& {Roberts}, B. 2002, \apj, 577, 475

\bibitem[{{Scharmer} {et~al.}(2008){Scharmer}, {Narayan}, {Hillberg}, {de la
  Cruz Rodr{\'{\i}}guez}, {L{\"o}fdahl}, {Kiselman}, {S{\"u}tterlin}, {van
  Noort}, \& {Lagg}}]{SCHetal2008}
{Scharmer}, G.~B., {Narayan}, G., {Hillberg}, T., {et~al.} 2008, \apjl, 689,
  L69

\bibitem[{{Terradas} {et~al.}(2011){Terradas}, {Arregui}, {Verth}, \&
  {Goossens}}]{TERetal2011}
{Terradas}, J., {Arregui}, I., {Verth}, G., \& {Goossens}, M. 2011, \apjl, 729,
  L22

\bibitem[{{Tripathi} {et~al.}(2012){Tripathi}, {Mason}, {Del Zanna}, \&
  {Bradshaw}}]{TRIetal2012}
{Tripathi}, D., {Mason}, H.~E., {Del Zanna}, G., \& {Bradshaw}, S. 2012, \apjl,
  754, L4

\bibitem[{{Ugarte-Urra} {et~al.}(2009){Ugarte-Urra}, {Warren}, \&
  {Brooks}}]{UGAetal2009}
{Ugarte-Urra}, I., {Warren}, H.~P., \& {Brooks}, D.~H. 2009, \apj, 695, 642

\bibitem[{{van Doorsselaere} {et~al.}(2007){van Doorsselaere}, {Nakariakov}, \&
  {Verwichte}}]{VANetal2007}
{van Doorsselaere}, T., {Nakariakov}, V.~M., \& {Verwichte}, E. 2007, \aap,
  473, 959

\bibitem[{{van Noort} {et~al.}(2005){van Noort}, {Rouppe van der Voort}, \&
  {L{\"o}fdahl}}]{VANNetal2005}
{van Noort}, M., {Rouppe van der Voort}, L., \& {L{\"o}fdahl}, M.~G. 2005,
  \solphys, 228, 191

\bibitem[{{Vaughan}(2005)}]{VAU2005}
{Vaughan}, S. 2005, \aap, 431, 391

\bibitem[{{Verth} {et~al.}(2008){Verth}, {Erd{\'e}lyi}, \&
  {Jess}}]{VERERDJES2008}
{Verth}, G., {Erd{\'e}lyi}, R., \& {Jess}, D.~B. 2008, \apjl, 687, L45

\bibitem[{{Verwichte} {et~al.}(2010){Verwichte}, {Foullon}, \& {Van
  Doorsselaere}}]{VERetal2010b}
{Verwichte}, E., {Foullon}, C., \& {Van Doorsselaere}, T. 2010, \apj, 717, 458

\bibitem[{{Verwichte} {et~al.}(2004){Verwichte}, {Nakariakov}, {Ofman}, \&
  {Deluca}}]{VERetal2004}
{Verwichte}, E., {Nakariakov}, V.~M., {Ofman}, L., \& {Deluca}, E.~E. 2004,
  \solphys, 223, 77

\bibitem[{{Viall} \& {Klimchuk}(2012)}]{VIAKLI2012}
{Viall}, N.~M. \& {Klimchuk}, J.~A. 2012, \apj, 753, 35

\bibitem[{{Wall} \& {Jenkins}(2003)}]{WALJEN}
{Wall}, J.~V. \& {Jenkins}, C.~R. 2003, Practical Statistics for Astronomers,
  Cambridge observing handbooks for research astronomers, vol.~3. (Cambridge,
  UK: Cambridge University Press)

\bibitem[{{Wang} {et~al.}(2015){Wang}, {Ofman}, {Sun}, {Provornikova}, \&
  {Davila}}]{WANetal2015}
{Wang}, T., {Ofman}, L., {Sun}, X., {Provornikova}, E., \& {Davila}, J.~M.
  2015, \apjl, 811, L13

\bibitem[{{White} {et~al.}(2013){White}, {Verwichte}, \&
  {Foullon}}]{WHIetal2013}
{White}, R.~S., {Verwichte}, E., \& {Foullon}, C. 2013, \apj, 774, 104

\bibitem[{{Winebarger} {et~al.}(2003){Winebarger}, {Warren}, \&
  {Seaton}}]{WINetal2003}
{Winebarger}, A.~R., {Warren}, H.~P., \& {Seaton}, D.~B. 2003, \apj, 593, 1164

\bibitem[{{Yuan} \& {Nakariakov}(2012)}]{YUANAK2012}
{Yuan}, D. \& {Nakariakov}, V.~M. 2012, \aap, 543, A9

\end{thebibliography}

\end{document}